\documentclass[a4paper,11pt]{article}

\usepackage[utf8]{inputenc}

\usepackage{gensymb} 

\usepackage{authblk} 
\usepackage{natbib} 

\usepackage{amsmath}
\usepackage{amssymb}
\usepackage{url}
\usepackage{hyperref}
\usepackage{caption} 
\usepackage{color,graphicx}
\usepackage{lineno}
\usepackage{verbatim}

\usepackage{xcolor}

\newcommand\hl[1]{%
  \bgroup
  \hskip0pt\color{red!80!black}%
  #1%
  \egroup
}

\newcommand{\Csharp}{%
  {\settoheight{\dimen0}{C}C\kern-.05em \resizebox{!}{\dimen0}{\raisebox{\depth}{\#}}}}

\newcommand\blfootnote[1]{%
  \begingroup
  \renewcommand\thefootnote{}\footnote{#1}%
  \addtocounter{footnote}{-1}%
  \endgroup
}

\begin{document}


\title{Testing SensoGraph, a geometric approach for fast sensory evaluation}

\author[1]{David Orden
}

\author[2]{Encarnaci\'on Fern\'andez-Fern\'andez
}

\author[3]{Jos\'e M. Rodr\'{\i}guez-Nogales
}

\author[4]{Josefina Vila-Crespo
}

\affil[1]{Departamento de F\'{i}sica y Matem\'{a}ticas,
    Universidad de Alcal\'{a}, Spain.
    \texttt{david.orden@uah.es}
}

\affil[2]{\'Area de Tecnolog\'{\i}a de Alimentos, E.T.S. de Ingenier\'{\i}as Agrarias, Universidad de Valladolid, Spain..
    \texttt{effernan@iaf.uva.es}, \texttt{rjosem@iaf.uva.es}, \texttt{jvila@pat.uva.es}
}

\date{}

\maketitle

\begin{abstract}
This paper introduces SensoGraph, a novel approach for fast sensory evaluation using two-dimensional geometric techniques. In the tasting sessions, the assessors follow their own criteria to place samples on a tablecloth, according to the similarity between samples. In order to analyse the data collected, first a geometric clustering is performed to each tablecloth, extracting connections between the samples.
Then, these connections are used to construct a global similarity matrix.
Finally, a graph drawing algorithm is used to obtain a 2D consensus graphic, which reflects the global opinion of the panel by (1) positioning closer those samples that have been globally perceived as similar and (2) showing the strength of the connections between samples. The proposal is validated by performing four tasting sessions, with three types of panels tasting different wines, and by developing a new software to implement the proposed techniques. The results obtained show that the graphics provide similar positionings of the samples as
the consensus maps obtained by multiple factor analysis (MFA), further providing extra information about connections between samples, not present in any previous method. The main conclusion is that the use of geometric techniques
provides information complementary to MFA,
and of a different type.
Finally, the method proposed is computationally able to manage a significantly larger number of assessors than MFA, which can be useful for the comparison of pictures by a huge number of consumers, via the Internet.
\end{abstract}


\blfootnote{\begin{minipage}[l]{0.2\textwidth} \includegraphics[trim=10cm 6cm 10cm 5cm,clip,scale=0.15]{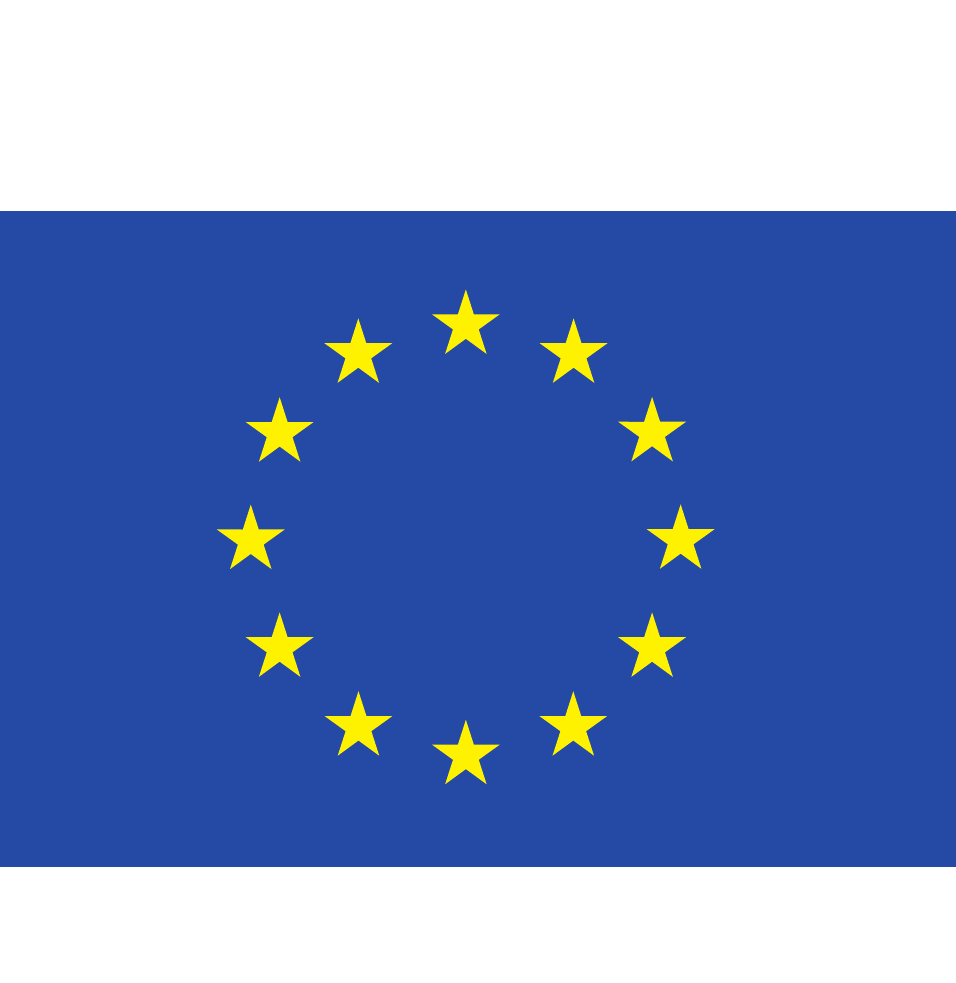} \end{minipage}  \hspace{-0.7cm}
\begin{minipage}[l][1cm]{0.8\textwidth}
 	  This work has received funding from the European Union's Horizon 2020 research and innovation programme under the Marie Sk\l{}odowska-Curie grant agreement No 734922.
 	\end{minipage}\vspace{-0.4cm}}



\section{Introduction and related work}
\label{sec:Introduction}

The aim of this work is to introduce and evaluate SensoGraph, a novel approach for the analysis of sensory data using geometric techniques
which deal with basic objects in 2D, like points, circles, and segments~\citep{Gabriel1969,Kamada1989}.
A data collection following the methodology introduced by~\citet{Risvik1994} for Projective Mapping is combined with a data analysis using geometric Multidimensional Scaling. A consensus graphic is obtained, showing not only a positioning of the samples, but also connections between samples and the force (strength) of these connections. This aims to be helpful in order to calibrate the significance of the positions on the graphic and to reflect the relations between groups. Moreover, the use of geometric techniques aims to help avoiding possible misuses of statistical techniques. The proposed method is validated by performing four sessions with three types of panels tasting different wines.

Sensory profiling is among the most important and widely used tools in sensory and consumer science~\citep{Lawless2010}, both in academia and industries~\citep{Varela2012}.
In these two fields, descriptive analysis has usefully linked product characteristics and consumer perception~\citep{Varela2012, Vidal2014}. Descriptive panels allow, due to their expertise, to obtain very detailed, robust, consistent, and reproducible results~\citep{Moussaoui2010}.
However, creating and maintaining a well-trained, calibrated, sensory panel can become too long and costly: For academic research, because of dealing with occasional projects or scarce funding~\citep{Lawless2010,Murray2001,Varela2012}. For companies, because of reasons like funding limits or difficulty to enrol assessors in a panel during a long time.

Thus, several alternative methods have arisen in the last years~\citep{Varela2012}, aiming to provide a fast sensory positioning of a set of products by assessors who are not necessarily trained. Skipping the need to train the panellists allows to elude the need of waiting a long time before obtaining results, as well as the need of agreeing on particular attributes, which may become difficult when working with experts like wine professionals or chefs~\citep{Hopfer2013}.
Introduced by~\cite{Risvik1994,Risvik1997}, Projective Mapping asks the assessors to position the presented samples on a two-dimensional space, usually a blank sheet of paper as tablecloth, following their own criteria: The more similar they perceive two samples, the closer they should position them, and vice versa~\citep{Perrin2008}. In those seminal works, the data were analysed by generalized procrustes analysis (GPA)~\citep{Gower1975} and principal component analysis (PCA)~\citep{Gabriel1971}, using the RV coefficient~\citep{Escoufier1979} to compare the method with conventional profiling.

More recently,~\citet{Pages2003,Pages2005a} proposed the use of multiple factor analysis (MFA)~\citep{Escofier1994} for data analysis, coining the name Napping\textsuperscript{\textregistered}. Typically, a two-dimensional graphic is obtained, where proximity of two samples indicates that the panel has globally perceived them to be similar.

The goal of these statistical methods is always to get an average configuration of products called consensus graphic, so it is crucial to assess its stability. Thus, in order to know whether two products are perceived as significantly different from a sensory point of view, the positions on the map given by these statistical methods should include a confidence area, e.g., confidence ellipses~\citep{Cadoret2013}.

Projective Mapping has been successfully used with many different kinds of products, among which the application to wine stands out~\citep{Piombino2004,Ballester2005,Pages2005a,Perrin2008,Perrin2009,Becue2011,Ross2012,Hopfer2013,Torri2013,Vidal2014}.
Other examples of
beverages analysed by these methods are
beers~\citep{Chollet2001,Abdi2007,Lelievre2008,Lelievre2009,Reinbach2014},
citrus juices~\citep{Nestrud2008},
drinking waters~\citep{Falahee1995,Falahee1997,Teillet2010}
high alcohol products~\citep{Louw2013},
hot beverages~\citep{Moussaoui2010},
lemon iced teas~\citep{Veinand2011},
powdered juices~\citep{Ares2011}, or
smoothies~\citep{Pages2010}.
The book by~\cite{Varela2014} details more products to which consumer based descriptive methodologies have been applied.

Until now, all the methodologies proposed for fast sensory evaluation have used statistical techniques to perform the data analysis. This paper introduces and evaluates a novel approach, a combination of geometric techniques to obtain a different kind of consensus graphic, here named SensoGraph. The outcome is a graph representation which combines a positioning of the samples together with connections representing the strength of the relations between them. Such a kind of representation is becoming more and more usual nowadays, among other reasons because of allowing dynamic data visualization~\citep{Beck2017taxonomy}, being helpful for big data visualization~\citep{Baumann2016twitter,Conover2011predicting,Junghanns2015gradoop}, and providing apparent graphics suitable for mass media~\citep{ViceNews2016} and the analysis of sports~\citep{buldu2018}.

\section{Material and methods}

\subsection{Data collection}\label{subsec:DataCollection}
In order to validate this proposal,
a total of four tasting sessions using Projective Mapping~\citep{Risvik1994,Pages2005a} have been performed, with three types of panels tasting different wines.

\textit{(A) Panel trained in Quantitative Descriptive Analysis (QDA):}
A panel trained in QDA of wine, composed of eleven assessors, tasted eight different red wines in one session, all of them elaborated at the winery of the School of Agricultural Engineering of the University of Valladolid in Palencia (Spain). Four of the wines were from cv.~Cabernet Sauvignon and the other four from cv.~Tempranillo, all of them from the same vintage. This panel was selected and trained using~\cite{ISO8586}.

\textit{(B) Panel receiving one training session in Projective Mapping:}
Another panel, composed of twelve assessors with experience in wine tasting, performed two sessions of Projective Mapping; a first session without any experience in the method and a repetition. The same eight red wines were used both for the training and for the final test, all of them elaborated at the winery of the School of Agricultural Engineering of the University of Valladolid in Palencia (Spain) using cv.~Tempranillo from Toro appellation (Spain)
and the same vintage. These eight wines were different from those tasted by the previous panel. This panel was composed by students of the Enology degree at the University of Valladolid, who had studied three academic years of Enology including a course in Sensory Analysis.

\textit{(C) Panel of habitual wine consumers tasting commercial wines:}
A final panel, composed of twenty-four habitual consumers of wine, performed one session of Projective Mapping. They tasted nine commercial wines, one of them duplicated. Seven of the wines used only one variety: Three of them were cv.~Menc\'{\i}a, three more were cv.~Tempranillo (one of them from Toro appellation, Spain), and another one was cv.~Monastrell. The other two wines were a blend of varieties: The duplicated wine used mainly cv.~Cabernet Franc, together with cv.~Merlot, Garnacha, and Monastrell. The other wine was mainly cv.~Tempranillo, blended with cv.~Garnacha and Graciano.

For all the sessions, the number of  samples followed the recommendations of~\cite{Valentin2016}. The samples were simultaneously presented to each assessor. The panellists were requested to position the wine samples on an A2 paper ($60\times 40$ cm), in such a way that two wine samples were to be placed close to each other if they seemed sensorially similar, and that two wines were to be distant from one another if they seemed sensorially different. All of this according to the assessor's own criteria for what close or far mean.

In all the sessions, the samples were served as 25 mL aliquots in standardised wineglasses~\citep{ISO3591}, which were coded with 3-digit numbers, and all the samples were presented simultaneously using a randomized complete block design. The serving temperature was 14$\pm 1\degree$C. All these sensory evaluations were carried out at the Sensory Science Laboratory of the School of Agricultural Engineering, at the University of Valladolid, Palencia (Spain), in individual booths designed in accordance with~\cite{ISO8589}.

\subsection{Data analysis}

The $x$- and $y$-coordinates of each sample on the paper were measured from the left-bottom corner of the sheet. These data were then stored in a table with $S$ rows, one for each sample, and $2A$ columns, with $A$ being the number of assessors.

\subsubsection{Statistical techniques}
On one hand, these data were analysed by statistical techniques with MFA, as proposed by~\cite{Pages2005a}, using the \texttt{R}~language~\citep{RDevelopmentCore2007} and the \texttt{FactoMineR} package~\citep{Le2008}.
MFA has become a common choice for the analysis of Projective Mapping data~\citep{Varela2014}, and it has been proved to be equal or better than other models like individual differences scaling (INDSCAL) for estimating the consensus configuration~\citep{naes2017}.
Finally, confidence  ellipses were constructed using truncated total bootstrapping~\citep{Cadoret2013} with \texttt{SensoMineR} package~\citep{sensominer}.

\subsubsection{Geometric techniques}
\label{subsec:steps}
On the other hand, in order to analyse the data by geometric techniques, we have developed and applied the following method:

\textbf{\underline{Step~1:}} Geometric clustering~\citep{capoyleas1991} allows to group data using basic operations from two-dimensional geometry, like drawing circles or segments. With the goal of analyzing each tablecloth to connections between the samples, and after exploring a large number of alternatives~\citep{SensoGraphEGC2013}, the Gabriel graph~\citep{Gabriel1969} was chosen, because of its good behavior and its clustering abilities having been widely checked~\citep{Matula1980,Urquhart1982,Choo2007}.

\begin{figure}[htb]
\begin{center}
\includegraphics[width=0.9\textwidth]{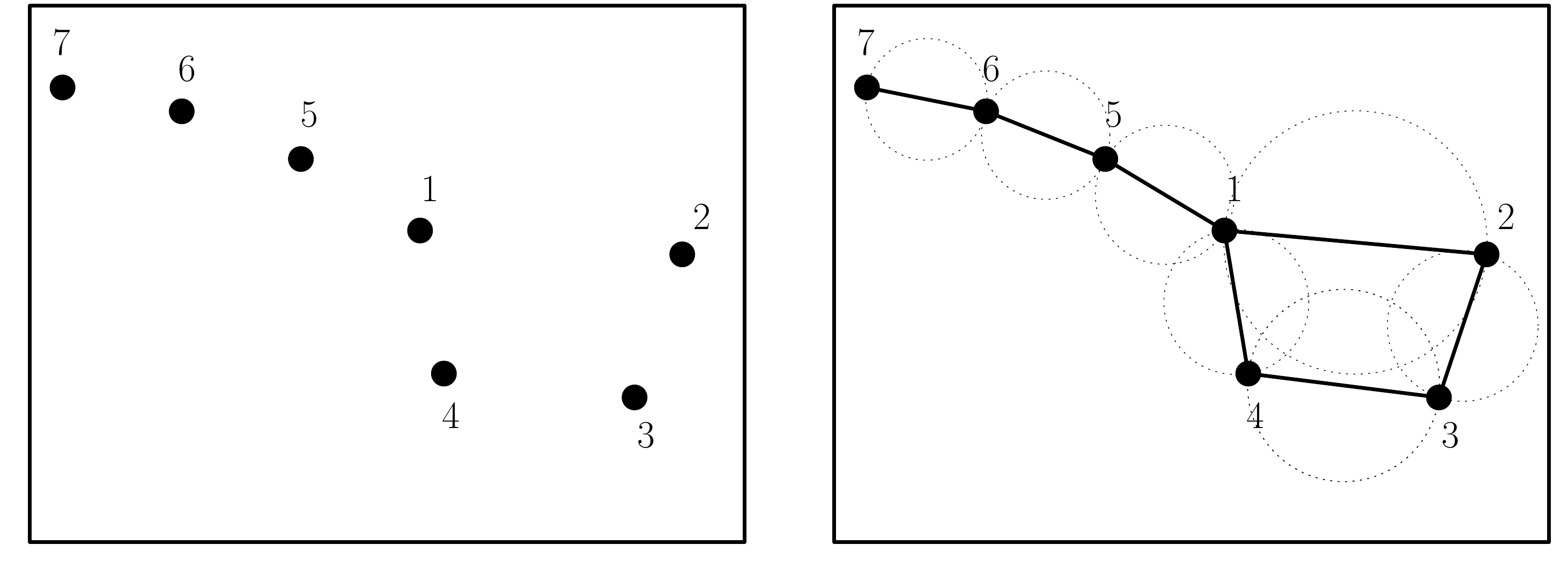}
\end{center}
\caption{
Left: Tablecloth positioning seven samples. Right: Construction of the associated Gabriel graph.}
\label{fig:Tablecloths}
\end{figure}

\begin{figure}[htb]
\begin{center}
\includegraphics[width=0.8\textwidth]{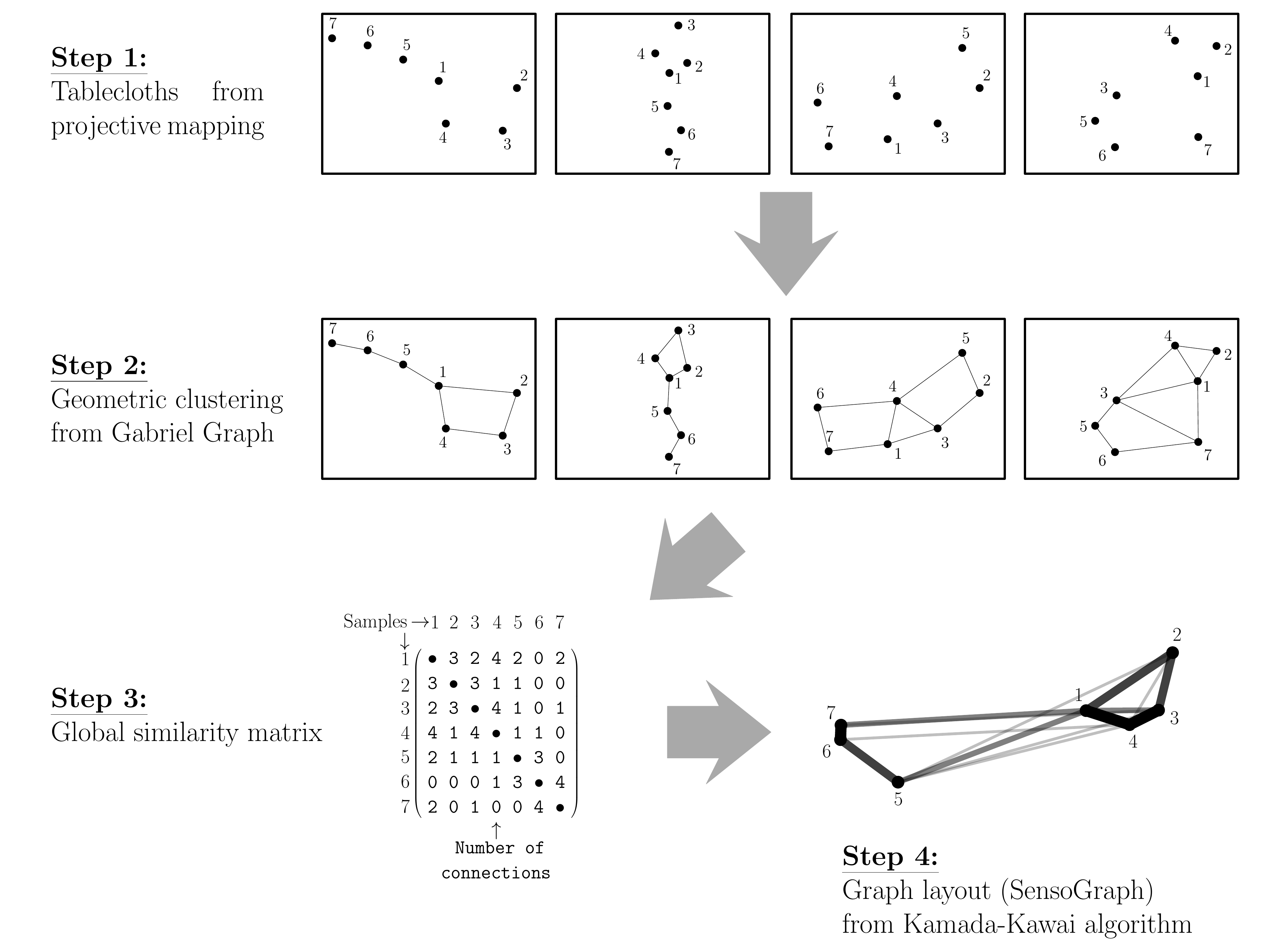}
\end{center}
\caption{First row: Four hypothetical tablecloths from Projective Mapping. Second row: Geometric clustering obtained by the Gabriel Graph. Third row, left: Global similarity matrix. For an example, entry $1,2$ has value~$3$ because the samples $1$ and~$2$ are connected in~$3$ of the tablecloths in the second row. Third row, right: The SensoGraph resulting graphic, where distances between samples represent the global similarity perceived. Two clear groups 1-2-3-4 and 5-6-7 appear, consistently with the tablecloths on top. In addition, connections between samples can be checked, with the connection forces being represented by thickness and opacity. For an example, the edge $1-2$, having force~$3$, is thicker and more opaque than the edge $2-5$, whose force is~$1$, but thinner and more transparent than the edge $1-4$, of force~$4$.}\label{fig:Similarities}
\end{figure}

For the construction of the Gabriel graph, two samples $P,Q$ get connected if, and only if, there is no other sample inside the closed disk having the straight segment $P-Q$ as diameter.
Figure~\ref{fig:Tablecloths} shows how to construct a Gabriel graph. Figure~\ref{fig:Similarities} shows another example, with four tablecloths (first row) and their corresponding Gabriel graphs (second row).

Recall that the assessors position the samples on the tablecloth without a common metric criterion, according to their own understanding of close and far. For an example, look at the two leftmost tablecloths in the top row of Figure~\ref{fig:Similarities}. The square 1-2-3-4 shows different distances in the two tablecloths, with samples 1-2 being much closer in the second picture than in the first one. However, at a glance we would say that both tablecloths provide similar information, namely a group 1-2-3-4 together with the samples 5-6-7 getting further from that group.

This is the kind of information extracted by the Gabriel graph, which therefore leads to the same graph for those two cases. See the two leftmost pictures of the second row in Figure~\ref{fig:Similarities}, which both show the same connections.

\textbf{\underline{Step~2:}} A global similarity matrix~\citep{Abdi2007}\label{matrix} was constructed. Each entry of the matrix stores, for a pair of samples $P,Q$, how many tablecloths show a connection $P-Q$ after the clustering step (e.g., entry $1,2$ will equal the number of tablecloths in which the samples $1$ and~$2$ are connected).
Figure~\ref{fig:Similarities} illustrates (third row, left) the global similarity matrix from four tablecloths for which the clustering Gabriel graph has already been constructed (second row).

This global similarity matrix can alternatively be seen as encoding a graph, in which entry $i,j$ stores the weight of the connection between vertices~$i$ and~$j$.

\textbf{\underline{Step~3:}}
A graph drawing algorithm~\citep{Eades2009} was applied to the graph encoded by the global similarity matrix.
Graph drawing algorithms have been used in social and behavioral sciences as a geometric alternative~\citep{deJordy2007} to non-metric multidimensional scaling~\citep{Chollet2014}. Among the different kinds of graph drawing algorithms, the particular class of force-directed drawing algorithms~\citep{Fruchterman1991,Hu2005} has been the one chosen, because of providing good results and being easy to understand.

In this class of algorithms, each entry $P,Q$ of the global similarity matrix
models the force of a spring, which connects $P-Q$ and pulls those samples together with that prescribed force.
The particular algorithm chosen has been the Kamada-Kawai algorithm, where the resulting system of forces is let to evolve until an equilibrium position of the samples is reached. Technical details can be checked at the paper by~\cite{Kamada1989}, but for a better understanding of this third step, the reader can imagine that the samples are (1)~pinned at arbitrary positions on a table, (2)~joined by springs with the forces specified in the matrix, and (3)~finally unpinned all at the same time so that they evolve to an equilibrium position.

Figure~\ref{fig:Similarities} shows a graphical sketch of these three steps. The equilibrium position reached provides a consensus graphic, here named SensoGraph, which reflects the global opinion of the panel by positioning closer those samples that have been globally perceived as similar. In addition, the graphic shows the connections and represents their forces by the thickness and opacity of the corresponding segments (the actual values of the forces being attached as a matrix). This information allows to know how similar or different two products have been perceived, playing the role of the confidence areas used by other methods in the literature~\citep{Cadoret2013}.

\subsection{Software}
\label{subsec:Software}

In order to perform the three steps detailed above, a new software was implemented. For convenience, Microsoft Visual Studio together with the programming language \Csharp\ were used to create an executable file for Windows, which allows to visually open the data spreadsheet and click a button to obtain the consensus graphic. This allows to start using the software with a negligible learning curve.

The implementation of Step~1 above followed a standard scheme for the construction of the Gabriel graph~\citep{Gabriel1969}, computing first the Delaunay triangulation~\citep{deBerg2008} and then traversing its edges to check which of them fulfill the Gabriel graph defining condition (that there is no other sample inside the closed disk having that edge as diameter, as stated in Step~1). Note that this is an exact algorithm, and hence there are no parameters to be chosen.

Implementing Step~2 was straightforward, just needing to run through the Gabriel graphs obtained, updating the counters for the appearances of each edge, and storing the results as a matrix.
Finally, for Step~3 the algorithm in the seminal paper by~\citet{Kamada1989} was used. This algorithm does need the following choices of parameters. The desirable length~$L$ of an edge in the outcome, for which the diameter of the tablecloths was used as suggested in Eq.~(3) in the reference. A constant $K=100$ used to define the strengths of the springs as in Eq.~(4) in the reference, which determines how strongly the edges tend to the desirable length. Finally, a maximum number $C=1000$ of iterations and a threshold $\varepsilon=0.1$ were chosen for the stopping condition of the algorithm. All these choices are rather standard, since our tests did not show huge variability among different choices.

A video showing the software in use has been broadcast~\citep{OrdenVideo2018} and readers interested in the software can contact the corresponding author. Moreover, the implementation of a Python version and an \texttt{R} package are projected for the future.

\section{Results and discussion}
\label{sec:results}

\subsection{Results and discussion by tasting session}

\begin{figure}[htb]
\begin{center}
\includegraphics[width=0.55\textwidth]{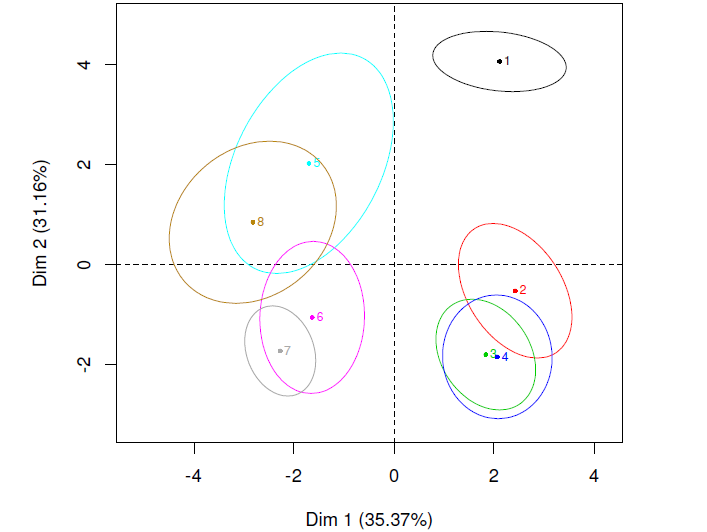}
\includegraphics[width=0.44\textwidth]{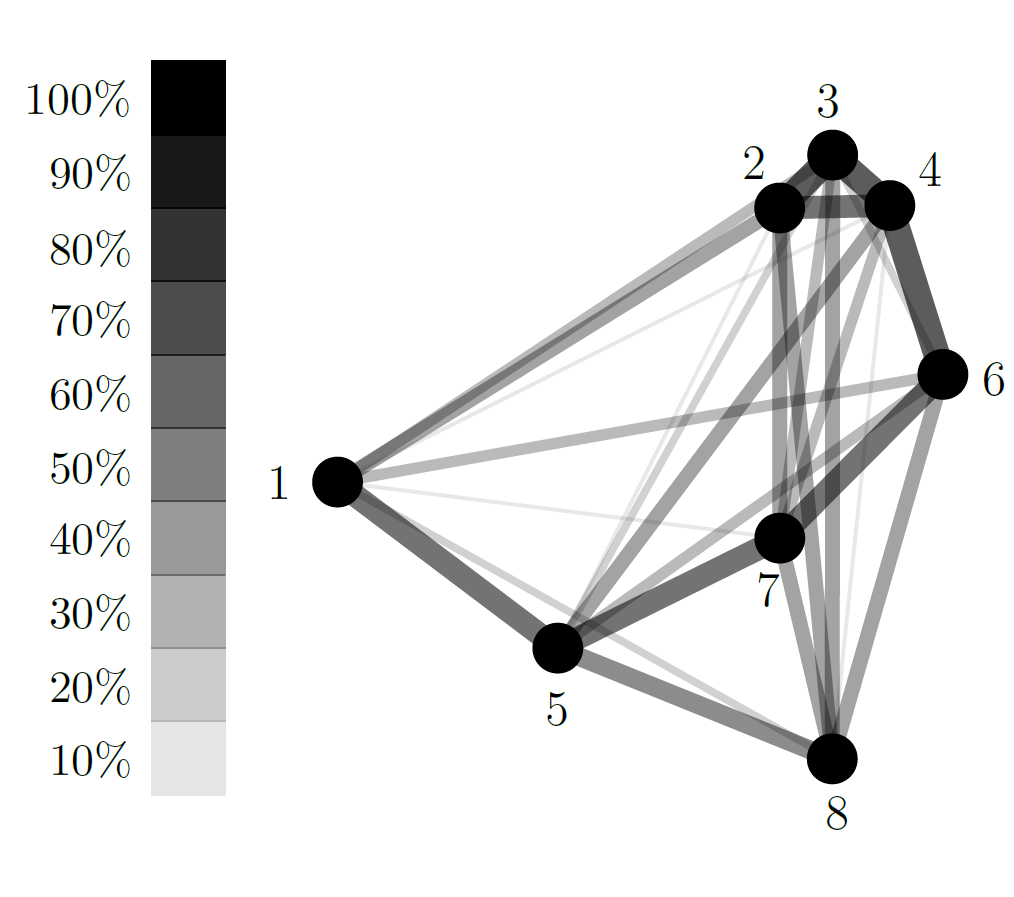}
\end{center}
\caption{Results for a first panel trained in QDA, composed by eleven assessors. Left: MFA. Right: SensoGraph and gradient for the percentage of tablecloths in which each connection arises.}\label{fig:Results1}
\end{figure}

\textit{(A) Panel trained in Quantitative Descriptive Analysis (QDA):}\\
The data obtained by performing Projective Mapping with the panel trained in QDA of wine were processed both by MFA (Fig.~\ref{fig:Results1}, left) and by SensoGraph (Fig.~\ref{fig:Results1}, right). For MFA, the first two dimensions accounted for 66.53\% of the explained variance.

At first glance, the positionings provided by MFA and SensoGraph look similar. Going into details, both graphics show a clear group 2-3-4: the corresponding ellipses in MFA superimpose, meaning that the assessors did not perceived a significant difference among these three samples, while in SensoGraph connections among 2-3-4 have arisen in between 55\% and 64\% of the tablecloths. (May the reader be interested in checking the actual numbers, these are shown in Table~\ref{table:Results1}.)

In addition, the graphic from MFA suggests a group 5-6-8, with non-empty intersection for their ellipses, and a further group 6-7. This connection 6-7 has appeared in the 55\% of the tablecloths in SensoGraph.
Further, in SensoGraph the connections in the group 5-6-8 have appeared in percentages from 27\% to 45\% of the tablecloths. On the contrary, the group 5-7-8 is more apparent, its connections having arisen in between the 36\% and the 55\% of the tablecloths.


This is because the geometric clustering has joined the sample 5 to the sample~7 in more tablecloths, 55\%, than to the sample~6, only 27\%.
It is interesting to note that this is compatible with the confidence ellipses of samples 5, 6, and~7 in the MFA graphic.

\begin{figure}[htb]
\begin{center}
\includegraphics[width=0.55\textwidth]{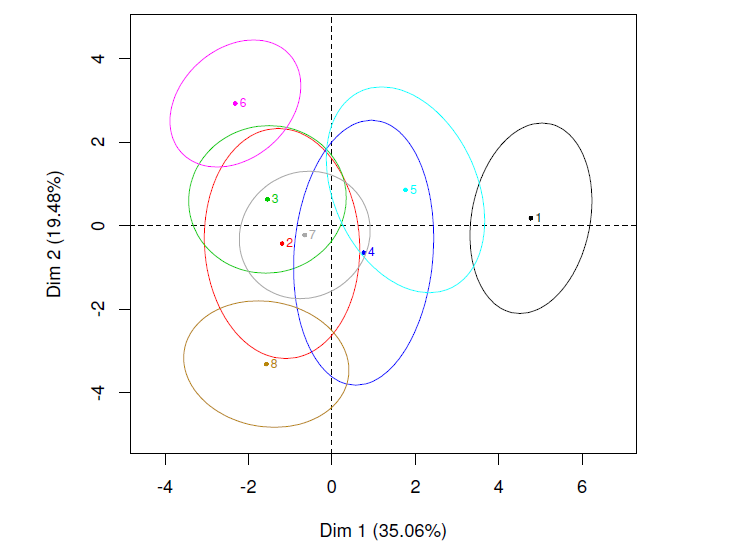}
\includegraphics[width=0.44\textwidth]{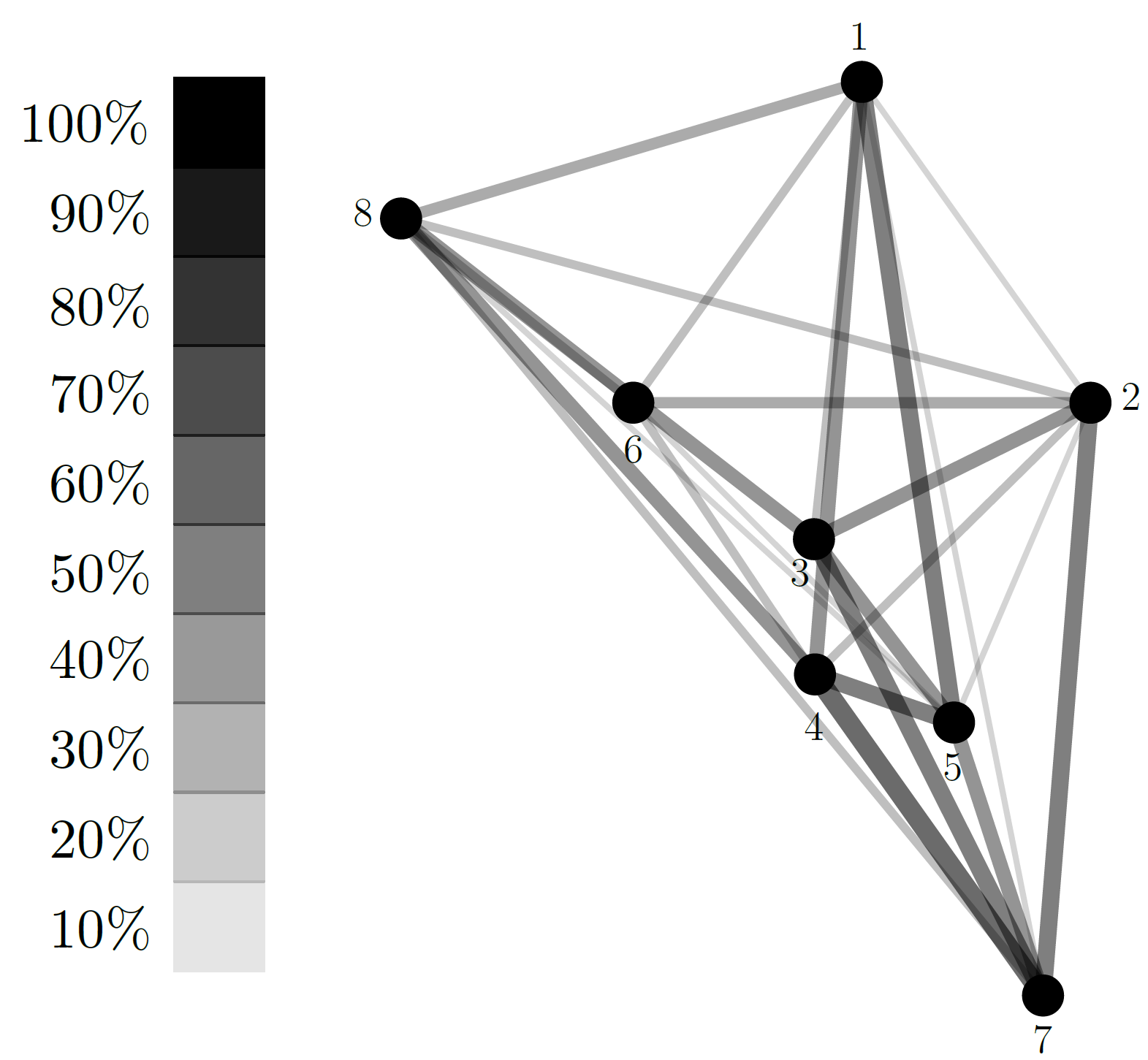}
\end{center}
\caption{Results for a panel without experience in the method, composed by twelve assessors. Left: MFA. Right: SensoGraph and gradient for the percentage of tablecloths in which each connection arises.}\label{fig:Results3}
\end{figure}

\begin{figure}[htb]
\begin{center}
\includegraphics[width=0.55\textwidth]{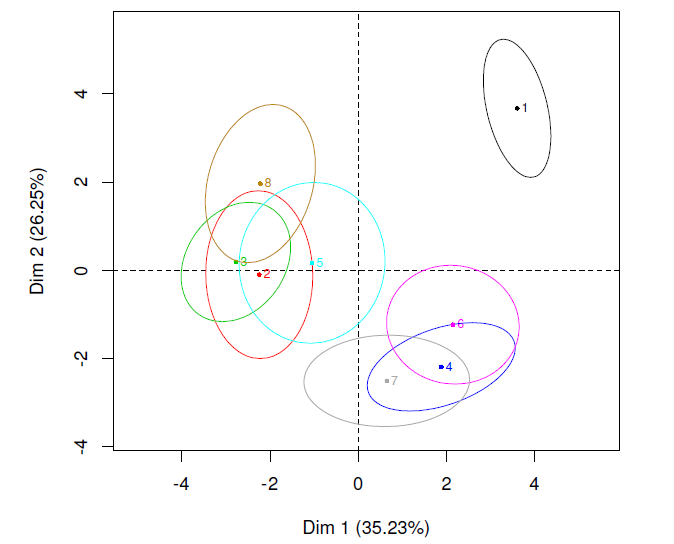}
\includegraphics[width=0.4\textwidth]{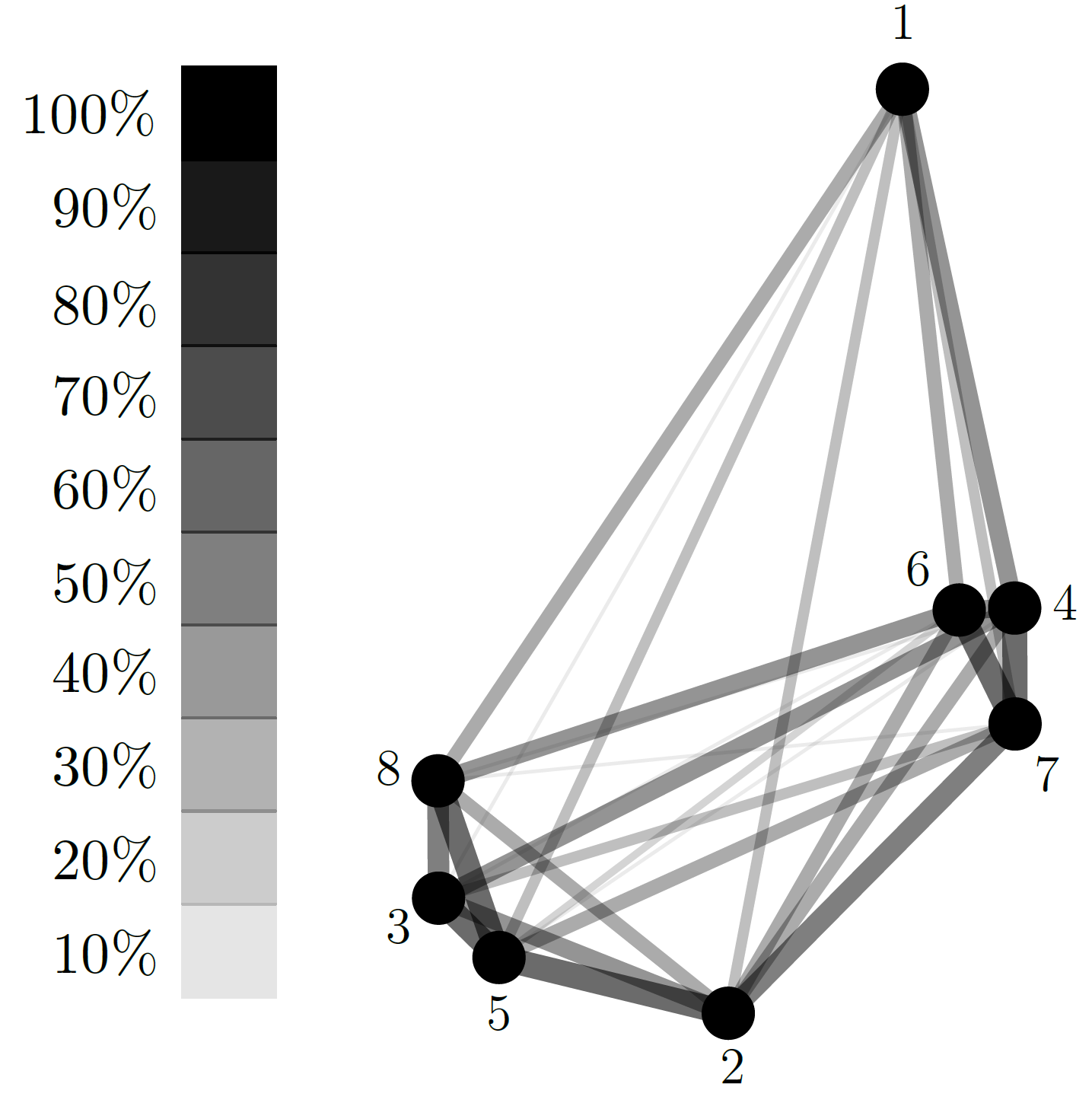}
\end{center}
\caption{Results for a second session with the panel in Fig.~\ref{fig:Results3}. Left: MFA. Right: SensoGraph and gradient for the percentage of tablecloths in which each connection arises.}\label{fig:Results4}
\end{figure}

\textit{(B) Panel receiving one training session in Projective Mapping:}\\
With the aim of studying how the experience in the Projective Mapping methodology affects the results, MFA and SensoGraph were used to process both the data obtained from a first Projective Mapping session with panellists having experience in tasting wines, Figure~\ref{fig:Results3}, and the data the same panel generated in a second session, Figure~\ref{fig:Results4}. For MFA, the first two dimensions accounted respectively for 54.54\% (Fig.~\ref{fig:Results3}, left) and 61.48\% (Fig.~\ref{fig:Results4}, left) of the explained variance, reflecting the effect of the experience achieved. For this panel, the comparison between MFA and SensoGraph positionings shows a higher coincidence when the percentage of explained variance is higher, i.e., in the second session.

The graphics in Figure~\ref{fig:Results3} are difficult to analyse even for a trained eye, since neither MFA nor SensoGraph show clear groups. The MFA plot shows the ellipses of samples 2-3-4-5-7 superimposed, while all of the ellipses of the remaining samples 1, 6, and~8 do, in turn, superimpose with some of the ellipses in the previous group.
In SensoGraph such a group of samples 2-3-4-5-7 appears indeed, at the lower-right corner, with connections ranging from the 8\% of tablecloths joining 3-4 to the 58\% joining 4-7.
Interestingly enough, SensoGraph allows to distinguish the behavior of sample~7, which turns out to be strongly connected to samples 2, 3, 4, and 5 in between the 42\% and the 58\% of the tablecloths,
from the behavior of sample~4, which is poorly connected with samples 2 and~3 since these connections arise, respectively, only at the 25\% and the 8\% of the tablecloths.

The situation for the repetition session is shown in Figure~\ref{fig:Results4}, where the same groups 2-3-5-8 and 4-6-7 are apparent for MFA (left) and SensoGraph (right), with sample~1 clearly isolated. In MFA the corresponding confidence ellipses do actually intersect, while in SensoGraph connections in the group 2-3-5-8 have appeared in
between the 33\% and the 58\% of the tablecloths
and those in the group 4-6-7 range 
between the 42\% and the 58\%.
Here, it is interesting to note
that sample~1 is better connected to the group 4-6-7,
these connections appearing in between 25\% and 42\% of the tablecloths, than to the group 2-3-5-8, 
in between 8\% and 33\%. (See Table~\ref{table:Results3-4} for the global similarity matrices.)

\textit{(C) Panel of habitual wine consumers tasting commercial wines:}\\

\begin{figure}[htb]
\begin{center}
\includegraphics[width=0.49\textwidth]{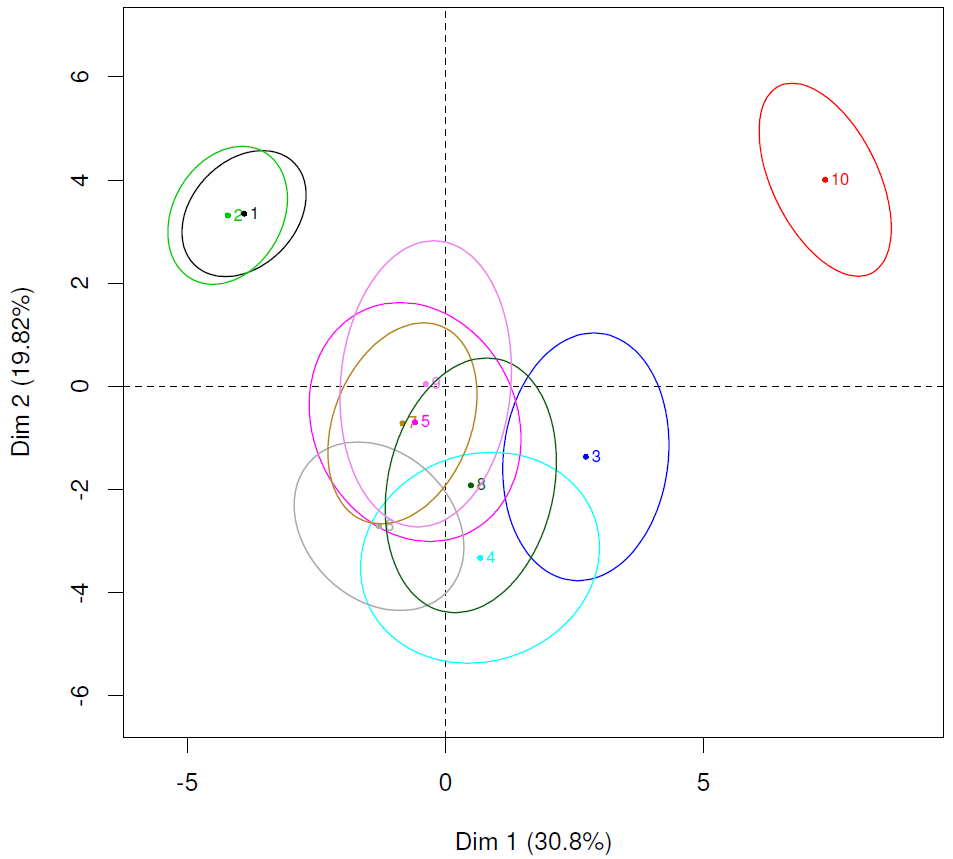}
\includegraphics[width=0.5\textwidth]{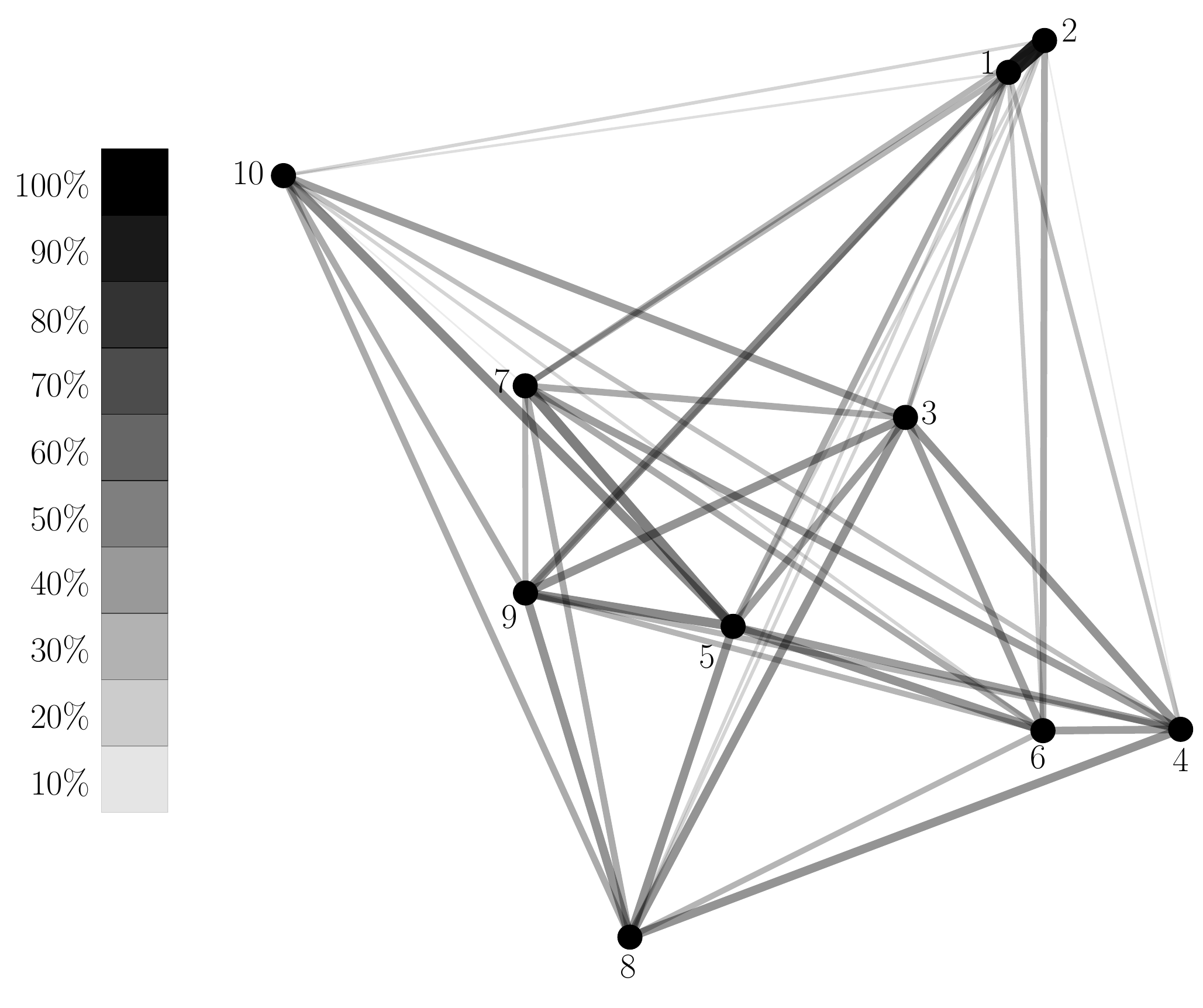}
\end{center}
\caption{Results for a panel composed by twenty-four habitual wine consumers tasting commercial wines. Left: MFA. Right: SensoGraph and gradient for the percentage of tablecloths in which each connection arises.}\label{fig:Results5}
\end{figure}

Again, the data were analyzed by MFA and SensoGraph, see Figure~\ref{fig:Results5}. For MFA, the first two dimensions accounted for 50.62\%. The positionings provided by between MFA and SensoGraph are similar, with samples~1-2 and~10 clearly separated from the others. This sample~10 used only cv.~Monastrell and the samples~1-2 correspond to wines elaborated with only cv.~Tempranillo.
It is interesting to observe that the pairs of samples 1-2 and 5-7 are quite similar in the MFA map, both according to the distances $d(1,2)$ and $d(5,7)$ between the two samples, and according to their corresponding ellipses. However, the pair 5-7 appears farther apart in SensoGraph, as can be checked at the global similarity matrix (Table~\ref{table:Results5}), where the connection 1-2 arises in an~88\% of the tablecloths, while the connection 5-7 arises in the~50\% of them. This is consistent with samples 1-2 being elaborated with only cv.~Tempranillo, while samples 5-7 differ in the grape variety, one of them using only cv.~Menc\'{\i}a and the other being a blend of cv.~Tempranillo, Garnacha and Graciano.

\subsection{General discussion}
As a summary of these four experiments, it can be observed that the positionings provided by SensoGraph are similar to those obtained by MFA.

Furthermore, the more trained is the panel,
the more clear and similar are the groups in the graphics given by SensoGraph and MFA. This is consistent with the behavior previously observed for statistical techniques, since \cite{Liu2016} reported that, for Projective Mapping,
conducting training on either the method or the product leads to more robust results.
Actually, observing the percentages of explained variance leads to the conclusion that, the higher the total inertia, the more similar are the positionings for MFA and SensoGraph.

Note that, for Step~3, it would be possible to use the Kamada-Kawai energy~\citep{gansner2004}, which is indeed analogous to the stress introduced by~\citet{kruskal1964multidimensional,kruskal1964nonmetric}, as an index of how well the graph drawing algorithm has drawn the data in the global similarity matrix. However, this would miss the effect of the geometric clustering in Step~1, for which an index of fit is not available.

In addition, it is interesting that the graphic for the SensoGraph method introduced in this paper does not provide only the positions for the samples, but also a graphical representation of the forces of connections, as well as a global similarity matrix. These connections and forces provide a better understanding of the interactions between groups, as already checked in different research fields~\citep{Beck2017taxonomy,Conover2011predicting,Junghanns2015gradoop}.
Further, these connections and forces help to calibrate the significance of the positioning~\citep{Cadoret2013}, with the help of the global similarity matrices.

For an example, they allow to contrast the distances in the map with the information of the tablecloths, as discussed in the last panel above. It is also interesting that for the matrix in Table~\ref{table:Results1} there is only one entry which is~\texttt{0}, while for those in Tables~\ref{table:Results3-4} and~\ref{table:Results5} there are no zero entries. This means that almost all samples have been connected at least once. Moreover, the connections appearing in the maximum number of tablecloths do so, respectively, in 58\%, 64\%, and 88\% of the tablecloths. These two observations show a large amount of individual variation in the data, which deserves further study.

Finally, concerning the usability of the SensoGraph method, on one hand the Projective Mapping methodology for data collection has already distinguished as natural and intuitive for the assessors~\citep{Ares2011,Varela2012,Carrillo2012}. On the other hand, the geometric techniques used for data analysis have been explained using basic geometric objects in~2D, aiming to be readily understood by researchers without any previous experience in the method.

\subsection{Computational efficiency}

Furthermore, all the geometric techniques used in this work are known to be extremely efficient from a computational point of view~\citep{Cardinal2009}.
In the following, the efficiency of the previous methods is analysed, using the standard \emph{big O notation} from algorithmic complexity. May the reader be unfamiliar with this notion, a good reference is, e.g., the book by~\citet{arora2009}. For the sake of an easier reading, a simplified explanation is also provided after the analysis.

First, the time complexity of SensoGraph is in~$O(AS\log(S)+AS+S^2)$, being $S$ the number of samples and $A$ the number of assessors as before. Each summand comes from each of the three steps detailed in Subsection~\ref{subsec:steps}, as follows: From the first step, the computation for each of the $A$ tablecloths of the Delaunay triangulation~\citep{deBerg2008} of $S$ samples, in~$O(S\log S)$, together with checking which of them does actually fulfill the condition to appear in the Gabriel graph. From the second step, counting the number of appearances of each of the $O(S)$ edges over the $A$ tablecloths. From the third step, the algorithm by~\citet{Kamada1989} takes $O(S^2)$ per each of the constant number of iterations stated in Subsection~\ref{subsec:Software}.

With the number $S$ of samples bounded in the order of tens, it is the number of assessors which can grow to the order of hundreds~\citep{Varela2014} or beyond. Hence, the complexity is dominated by the number~$A$ of assessors, and therefore the time complexity of SensoGraph is in~$O(A)$, i.e., linear in the number of assessors. On the contrary, the time complexity of~MFA is in~$O(A^3)$, i.e.,
cubic in the number of assessors, since it needs two rounds of~PCA~\citep{abdi2013}, whose time complexity is cubic~\citep{feng2000}.

\begin{figure}[htb]
\begin{center}
\includegraphics[width=0.49\textwidth]{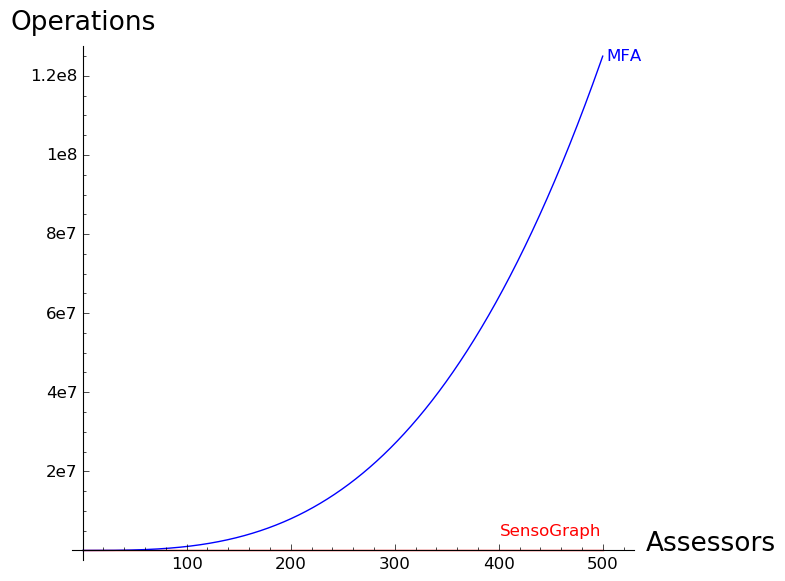}
\includegraphics[width=0.49\textwidth]{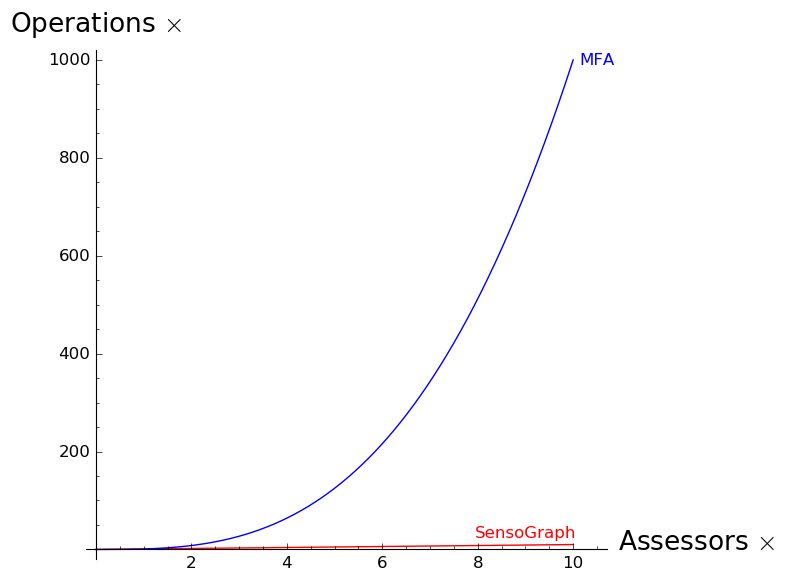}
\end{center}
\caption{
Left: Number of operations needed for a number of assessors up to~500. Right: Factor by which the number of operations gets multiplied when increasing by a factor the number of assessors (independently of the actual number of assessors considered).}
\label{fig:Complexities}
\end{figure}

An explanation in short of these two complexities is the following: Multiplying the number of assessors by a factor~$X$, the number of operations needed by SensoGraph gets multiplied by~$X$ as well, while the one needed by MFA gets multiplied by~$X^3$. For an example, duplicating the number of assessors (i.e., $X=2$), the number of operations needed by SensoGraph gets duplicated as well, while that of MFA gets multiplied by~$2^3$. For another example, if the number of assessors gets multiplied by~$10$, so does the number of operations needed by SensoGraph, while the number of operations needed by MFA gets multiplied by~$10^3$. The difference between these two growing rates is small for a number of assessors around~100, but apparent already for~200 and crucial when intending to work with a larger number of assessors, see Figure~\ref{fig:Complexities}.

Working with a greater number of assessors is likely to become more relevant, since sensory analysis moves towards the use of untrained consumers to evaluate products~\citep{Valentin2016}. Thanks to its linear time complexity, SensoGraph would be able to handle even millions of tablecloths~\citep{deBerg2008}, and this opens an interesting door towards massive sensory analysis, using the Internet to collect large datasets~\citep{Beck2017taxonomy,Conover2011predicting,Junghanns2015gradoop}. This feature can be particularly suitable for the comparison of pictures like, e.g., the one performed by~\cite{mielby2014comparison}. The use of photographs as surrogates of samples has been suggested by~\cite{maughan2016procedure} after proper validation of the photographs.


\section{Conclusions}

The main conclusion is that the use of geometric techniques can be an interesting complement to the use of statistics.
SensoGraph does not aim to substitute the use of statistics for the analysis of Projective Mapping data, but to provide an additional point of view for an enriched vision.

The results obtained by SensoGraph are comparable to those given by
the consensus maps obtained by MFA,
further providing information about the connections between samples. This extra information, not provided by any of the previous methods in the literature, helps to a better understanding of the relations inside and between groups.

In addition, we obtain a global similarity matrix storing the information about how many tablecloths show a connection between two samples. This is useful, for instance, when in the MFA map the distance $d(P_1,Q_1)$ between a pair of samples~$P_1,Q_1$ is very similar to the distance $d(P_2,Q_2)$ between a different pair of samples~$P_2,Q_2$. Comparing the two entries in the global similarity matrix allows to check whether the connections~$P_1-Q_1$ and~$P_2-Q_2$ do actually arise in a similar number of tablecloths or not.

Finally, the time complexity of SensoGraph is significantly lower than that of~MFA. This allows to efficiently manage a number of tablecloths several orders of magnitude above the one handled by MFA. This feature is of particular interest, provided the increasing importance of consumers for the evaluation of existing and new products, opening a door to the analysis of massive sensory data. A good example is the comparison of pictures as surrogates of samples, via the Internet, by a huge amount of assessors.


\section{Acknowledgements}

The authors want to gratefully thank professor Ferran Hurtado, in memoriam, for suggesting that proximity graphs could be used for the analysis of tablecloths. They also want to thank David N. de Miguel and Lucas Fox for implementing in a software the methods used.
We are thankful for the very helpful comments and input from two anonymous reviewers.

All the authors have been supported by the University of Alcal\'a grant CCGP2017-EXP/015. In addition, David Orden has been partially supported by MINECO Projects MTM2014-54207 and MTM2017-83750-P, as well as by H2020-MSCA-RISE project 734922 - CONNECT.

\section*{Tables}

\begin{table}[htb]
    \centering
{\small
$
\left(
\begin{array}{cccccccc}
\bullet & \texttt{4} & \texttt{3} & \texttt{1} & \texttt{6} & \texttt{3} & \texttt{1} & \texttt{2} \\
\texttt{4} & \bullet & \texttt{7} & \texttt{6} & \texttt{1} & \texttt{0} & \texttt{4} & \texttt{4} \\
\texttt{3} & \texttt{7} & \bullet & \texttt{7} & \texttt{2} & \texttt{2} & \texttt{3} & \texttt{4} \\
\texttt{1} & \texttt{6} & \texttt{7} & \bullet & \texttt{4} & \texttt{7} & \texttt{3} & \texttt{1} \\
\texttt{6} & \texttt{1} & \texttt{2} & \texttt{4} & \bullet & \texttt{3} & \texttt{6} & \texttt{5} \\
\texttt{3} & \texttt{0} & \texttt{2} & \texttt{7} & \texttt{3} & \bullet & \texttt{6} & \texttt{4} \\
\texttt{1} & \texttt{4} & \texttt{3} & \texttt{3} & \texttt{6} & \texttt{6} & \bullet & \texttt{4} \\
\texttt{2} & \texttt{4} & \texttt{4} & \texttt{1} & \texttt{5} & \texttt{4} & \texttt{4} & \bullet \\
\end{array}
\right)
$
}
\caption{Global similarity matrix for Figure~\ref{fig:Results1}.}
\label{table:Results1}
\end{table}

\begin{table}[htb]
    \centering
{\small
$
\left(
\begin{array}{cccccccc}
\bullet & \texttt{2} & \texttt{3} & \texttt{5} & \texttt{6} & \texttt{3} & \texttt{2} & \texttt{4} \\
\texttt{2} & \bullet & \texttt{5} & \texttt{3} & \texttt{2} & \texttt{4} & \texttt{6} & \texttt{3} \\
\texttt{3} & \texttt{5} & \bullet & \texttt{1} & \texttt{5} & \texttt{5} & \texttt{6} & \texttt{5} \\
\texttt{5} & \texttt{3} & \texttt{1} & \bullet & \texttt{6} & \texttt{3} & \texttt{7} & \texttt{5} \\
\texttt{6} & \texttt{2} & \texttt{5} & \texttt{6} & \bullet & \texttt{2} & \texttt{5} & \texttt{2} \\
\texttt{3} & \texttt{4} & \texttt{5} & \texttt{3} & \texttt{2} & \bullet & \texttt{3} & \texttt{3} \\
\texttt{2} & \texttt{6} & \texttt{6} & \texttt{7} & \texttt{5} & \texttt{3} & \bullet & \texttt{3} \\
\texttt{4} & \texttt{3} & \texttt{5} & \texttt{5} & \texttt{2} & \texttt{3} & \texttt{3} & \bullet \\
\end{array}
\right)
$\\
\smallskip

$
\left(
\begin{array}{cccccccc}
\bullet & \texttt{3} & \texttt{1} & \texttt{5} & \texttt{3} & \texttt{4} & \texttt{3} & \texttt{4} \\
\texttt{3} & \bullet & \texttt{5} & \texttt{5} & \texttt{7} & \texttt{4} & \texttt{6} & \texttt{4} \\
\texttt{1} & \texttt{5} & \bullet & \texttt{5} & \texttt{7} & \texttt{1} & \texttt{3} & \texttt{6} \\
\texttt{5} & \texttt{5} & \texttt{5} & \bullet & \texttt{1} & \texttt{5} & \texttt{7} & \texttt{1} \\
\texttt{3} & \texttt{7} & \texttt{7} & \texttt{1} & \bullet & \texttt{2} & \texttt{4} & \texttt{7} \\
\texttt{4} & \texttt{4} & \texttt{1} & \texttt{5} & \texttt{2} & \bullet & \texttt{7} & \texttt{5} \\
\texttt{3} & \texttt{6} & \texttt{3} & \texttt{7} & \texttt{4} & \texttt{7} & \bullet & \texttt{1} \\
\texttt{4} & \texttt{4} & \texttt{6} & \texttt{1} & \texttt{7} & \texttt{5} & \texttt{1} & \bullet \\
\end{array}
\right)
$
}
\caption{Global similarity matrices for Figures~\ref{fig:Results3} (top) and~\ref{fig:Results4} (bottom).}
\label{table:Results3-4}
\end{table}

\begin{table}[htb]
    \centering
{\small
$
\left(
\begin{array}{cccccccccc}
\bullet & \texttt{21} & \texttt{6} & \texttt{6} & \texttt{8} & \texttt{5} & \texttt{7} & \texttt{4}  & \texttt{11} & \texttt{3} \\
\texttt{21} & \bullet & \texttt{5} & \texttt{2} & \texttt{4} & \texttt{8} & \texttt{7} & \texttt{4}  & \texttt{5} & \texttt{4} \\
\texttt{6} & \texttt{5} & \bullet & \texttt{10} & \texttt{9} & \texttt{9} & \texttt{8} & \texttt{10}  & \texttt{10} & \texttt{9} \\
\texttt{6} & \texttt{2} & \texttt{10} & \bullet & \texttt{9} & \texttt{9} & \texttt{9} & \texttt{10}  & \texttt{7} & \texttt{6} \\
\texttt{8} & \texttt{4} & \texttt{9} & \texttt{9} & \bullet & \texttt{10} & \texttt{12} & \texttt{10}  & \texttt{11} & \texttt{7} \\
\texttt{5} & \texttt{8} & \texttt{9} & \texttt{9} & \texttt{10} & \bullet & \texttt{8} & \texttt{7}  & \texttt{7} & \texttt{4} \\
\texttt{7} & \texttt{7} & \texttt{8} & \texttt{9} & \texttt{12} & \texttt{8} & \bullet & \texttt{8}  & \texttt{7} & \texttt{2} \\
\texttt{4} & \texttt{4} & \texttt{10} & \texttt{10} & \texttt{10} & \texttt{7} & \texttt{8}  & \bullet & \texttt{10} & \texttt{8} \\
\texttt{11} & \texttt{5} & \texttt{10} & \texttt{7} & \texttt{11} & \texttt{7} & \texttt{7}  & \texttt{10} & \bullet & \texttt{8} \\
\texttt{3} & \texttt{4} & \texttt{9} & \texttt{6} & \texttt{7} & \texttt{4} & \texttt{2}  & \texttt{8} & \texttt{8} & \bullet \\
\end{array}
\right)
$
}
\caption{Global similarity matrix for Figure~\ref{fig:Results5}.}
\label{table:Results5}
\end{table}

\bibliographystyle{apalike}

\end{document}